\documentclass[9pt,conference]{IEEEtran}

\usepackage[preprint]{waspaa25}

\usepackage{bm} 
 \usepackage{multirow}
\usepackage{enumitem}

\title{Deciphering GunType Hierarchy through Acoustic Analysis of Gunshot Recordings}

\name{Ankit Shah$^{1}$,
      Rita Singh$^{1}$,
      Bhiksha Raj$^{1}$,
      Alexander Hauptmann$^{1}$}
\address{$^{1}$ Carnegie Mellon University, Pittsburgh PA, USA \;
}

\begin{document}

\maketitle

\begin{abstract}

The escalating rates of gun-related violence and mass shootings represent a significant threat to public safety. Timely and accurate information for law enforcement agencies is crucial in mitigating these incidents. Current commercial gunshot detection systems, while effective, often come with prohibitive costs. This research explores a cost-effective alternative by leveraging acoustic analysis of gunshot recordings, potentially obtainable from ubiquitous devices like cell phones, to not only detect gunshots but also classify the type of firearm used. This paper details a study on deciphering gun type hierarchies using a curated dataset of 3459 recordings. We investigate the fundamental acoustic characteristics of gunshots, including muzzle blasts and shockwaves, which vary based on firearm type, ammunition, and shooting direction. We propose and evaluate machine learning frameworks, including Support Vector Machines (SVMs) as a baseline and a more advanced Convolutional Neural Network (CNN) architecture for joint gunshot detection and gun type classification. Results indicate that our deep learning approach achieves a mean average precision (mAP) of 0.58 on clean labeled data, outperforming the SVM baseline (mAP 0.39). Challenges related to data quality, environmental noise, and the generalization capabilities when using noisy web-sourced data (mAP 0.35) are also discussed. The long-term vision is to develop a highly accurate, real-time system deployable on common recording devices, significantly reducing detection costs and providing critical intelligence to first responders.

\end{abstract}

\section{Introduction}
\label{sec:intro}

 Gun violence remains a critical issue affecting public safety globally, with particularly alarming statistics in regions like the United States. In the United States alone, firearms were responsible for 48,830 deaths in 2021, representing 1.6\% of total deaths in the USA, and a staggering 271 instances of mass shootings resulted in 1958 fatalities ~\cite{gunviolencearchive} between 2006 and 2025. These figures underscore the urgent need for effective tools and strategies to aid law enforcement in responding to and potentially preventing such incidents.

 Current gunshot detection systems, such as ShotSpotter~\cite{soundthinking2022}, demonstrate 97\% accuracy in controlled environments but require significant infrastructure investment—approximately \$65,000 per square mile annually for deployment and maintenance. This cost barrier prevents adoption in many communities that need protection most. We propose leveraging the computational capabilities of ubiquitous mobile devices to create a distributed, low-cost alternative. The core idea is to detect gunshots and, crucially, identify the type of firearm used from audio recordings captured by these devices. By analyzing acoustic signatures from smartphones and other common recording devices, our system can detect gunshots and classify firearm types (handgun, rifle, shotgun, etc.), providing critical intelligence for tactical response planning~\cite{millet2006latest}.


However, developing a robust gun type detection system presents several challenges. \cite{djeddou2013classification} Firstly, acquiring clean, accurately labeled data for various gun types is difficult. Many existing annotated datasets may feature a limited variety of gun types or suffer from a lack of diversity in recording conditions \cite{kiktova2015gun}. Secondly, the quality of recordings captured in real-world scenarios is often compromised by environmental noise (e.g., traffic, voices, echoes), varying microphone characteristics, and distance from the sound source. \cite{raponi2022sound} These factors can significantly degrade the distinctive acoustic signatures of gunshots, making classification a complex task. \cite{singh2022measurements}

This paper addresses these challenges by:
\begin{enumerate}
    \item Curating a comprehensive dataset of gunshot recordings across five broad gun type categories.
    \item Analyzing the fundamental acoustic properties of gunshots that differentiate various firearm types.
    \item Proposing and evaluating machine learning models, including a deep learning architecture, for the joint task of gunshot detection and gun type classification.
    \item Investigating the performance of these models on both clean annotated data and more challenging, noisy data sourced from the web.
\end{enumerate}
The ultimate aim is to lay the groundwork for a system that can significantly reduce the cost of gunshot detection and classification, making such technology more widely available and thereby enhancing public safety.

\section{Fundamentals of Gunshot Acoustics}
\label{sec:acoustics}
Understanding the acoustic phenomena associated with a gunshot is paramount for developing effective detection and classification systems. A typical gunshot sound comprises two primary components: the muzzle blast and, for supersonic projectiles, a ballistic shockwave (also known as a sonic boom) \cite{Maher2007AcousticalCO, Maher2008DecipheringGR}.

The \textbf{muzzle blast} is a high-intensity, short-duration acoustic pulse generated by the rapid expansion of propellant gases exiting the firearm's muzzle. This event typically lasts for about 3 to 5 milliseconds. Its characteristics are influenced by the type and amount of gunpowder, the barrel length, and the presence of any muzzle devices (e.g., suppressors, flash hiders). \cite{Maher2007AcousticalCO} The \textbf{shockwave} is produced by a projectile traveling at supersonic speeds (faster than the speed of sound). As the projectile moves through the air, it creates a conical pressure wave. The shock wave is a very brief event that often lasts only 200 to 400 microseconds. Its presence and characteristics depend on the velocity, shape, and size of the projectile, as well as its trajectory relative to the microphone. \cite{Maher2007AcousticalCO, 10059143}

The overall acoustic signature of a shot is a complex interplay of these components and is further shaped by environmental factors such as reflections, reverberation, and ambient noise. The key factors influencing the recorded sound include \cite{Maher2007AcousticalCO, Maher2008DecipheringGR, liang2019technical}:
\begin{itemize}[leftmargin=*]
    \item \textbf{Type and size of the firearm:} Different firearm mechanisms (e.g. revolver, semiautomatic pistol, bolt action rifle, shotgun) and calibers produce distinct muzzle blast profiles and, if applicable, shock wave characteristics. For instance, rifles generally produce much higher muzzle energy and projectile velocities than handguns.
    \item \textbf{Characteristics of the ammunition:} The type of propellant, weight of the bullet, and the design contribute to the sound. Subsonic ammunition, for example, will not produce a ballistic shockwave.
    \item \textbf{Direction of the gunshot relative to the microphone/observer and barrel orientation:} The perceived sound, particularly the timing and amplitude relationship between the muzzle blast and shockwave, changes significantly with the listener's position relative to the firearm's muzzle and the bullet's trajectory. \cite{maher2006modeling}
\end{itemize}

Table~\ref{tab:acoustic_characteristics} summarizes the key acoustic differences between firearm categories based on our analysis and prior literature~\cite{Maher2007AcousticalCO}.

\begin{table}[h!]
  \centering
  \caption{Typical Acoustic Characteristics by Firearm Type}
  \label{tab:acoustic_characteristics}
  \resizebox{\columnwidth}{!}{%
  \begin{tabular}{@{}lcccc@{}}
    \toprule
    \textbf{Firearm Type} & \textbf{Peak Freq (Hz)} & \textbf{Duration (ms)} & \textbf{SPL at 1m (dB)} & \textbf{Shockwave} \\
    \midrule
    Handgun (9mm)    & 500-2000  & 3-5   & 159-164 & Rare \\
    Rifle (.223)     & 200-1500  & 5-8   & 167-171 & Common \\
    Shotgun (12ga)   & 100-800   & 8-12  & 161-165 & No \\
    Machine Gun      & 300-1800  & 3-5*  & 165-170 & Common \\
    Submachine Gun   & 400-2200  & 3-4*  & 160-166 & Variable \\
    \bottomrule
  \end{tabular}%
  }
  \footnotesize{*Per shot in burst mode}
\end{table}



\begin{figure}[h!]
  \centering
  \includegraphics[width=0.9\linewidth]{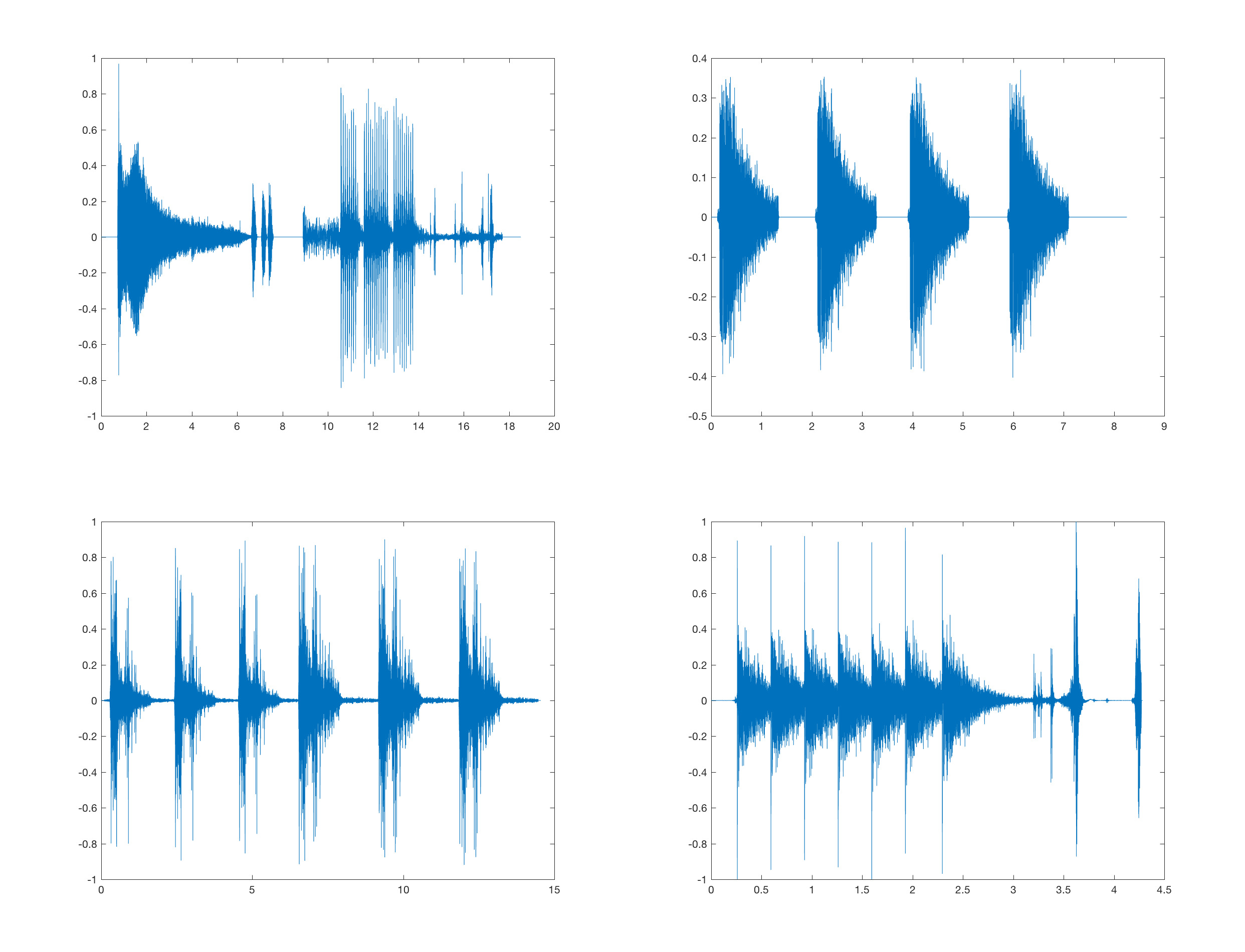} 
  \caption{Waveform examples for different gun types: a) AK-47 Gunshot Waveforms, b) M9 Pistol Waveform, c) Sniper Waveform, and d) Shotgun waveform. These illustrate that humans can often identify the broad category of gun type from visual inspection of waveforms.}
  \label{fig:gun_waveforms}
  \vspace{-2mm}
\end{figure}

Further visual inspection of waveforms from different gun categories, as shown in Figure \ref{fig:gun_waveforms}, suggests that even humans can often discern broad categories of gun types based on these visual representations. For instance, the rapid succession of shots from an automatic weapon like an AK-47 (Figure \ref{fig:gun_waveforms}a) differs significantly from the single, distinct report of a shotgun (Figure \ref{fig:gun_waveforms}d) or a sniper rifle (Figure \ref{fig:gun_waveforms}c). Pistols (Figure \ref{fig:gun_waveforms}b) often have a sharper, quicker report compared to some rifles. These observable differences form the basis for automated classification using machine learning techniques.

\section{Dataset Acquisition and Preparation}
\label{sec:dataset}
A critical component for training and evaluating any machine learning model for gun type classification is a comprehensive and well-annotated dataset. Due to the scarcity of publicly available, large-scale, labeled gunshot audio datasets specifically categorized by gun type hierarchy, we undertook the task of curating our own.

Recordings were collected from a variety of sources to ensure diversity:
\begin{itemize}[leftmargin=*]
    \item \textbf{CMU Database:} Internal audio collections at Carnegie Mellon University.
    \item \textbf{Spotify Playlists:} Publicly accessible playlists containing sound effects, including gunshots.
    \item \textbf{Airborne Sound (Firearm Sound Effects Library):} Commercially available sound effect libraries known for their high-quality recordings.
    \item \textbf{Web data}: While explored for broader data collection, this paper focuses on the curated and labeled dataset for primary evaluation. 
\end{itemize}

The collected audio clips were manually reviewed and annotated into five broad firearm categories: Rifle, Submachine Gun, Handgun/Pistol, Machine Gun, and Shotgun. These categories represent a practical hierarchy that balances specificity with the feasibility of acoustic differentiation. The distribution of recordings across these classes and their total duration are summarized in Table \ref{tab:dataset_summary}.

\begin{table}[h!]
  \centering
  \caption{Summary of the Curated Gunshot Recording Dataset.}
  \label{tab:dataset_summary}
  \resizebox{0.85\columnwidth}{!}{%
  \begin{tabular}{@{}lcc@{}}
    \toprule
    \textbf{Class} & \textbf{Number of Recordings} & \textbf{Total Duration (hrs)} \\
    \midrule
    Rifle          & 892                           & 1.67                    \\
    Submachine Gun & 522                           & 1.34                    \\
    Handgun/Pistol & 1105                          & 1.88                    \\
    Machine Gun    & 543                           & 2.21                    \\
    Shotgun        & 396                           & 1.32                    \\
    \midrule
    \textbf{Total} & \textbf{3459}                 & \textbf{8.4}            \\
    \bottomrule
  \end{tabular}%
  }
\end{table}

The dataset comprises a total of 3459 recordings, amounting to 8.4 hours of audio. While efforts were made to balance the classes, some inherent imbalance exists, with Handgun/Pistol being the most represented class and Shotgun the least. This distribution reflects the relative availability of clear recordings for these categories from the sourced materials.

\section{Experimental Setup and Methodology}
\label{sec:methodology}
This section details the audio feature extraction process, the machine learning classifiers employed, and the experimental design used for evaluating the proposed gun type classification system.

\subsection{Audio Feature Extraction}
\label{ssec:features}
To prepare the audio data for machine learning models, all recordings underwent a standardized preprocessing pipeline:
\begin{enumerate}
    \item \textbf{Resampling and Channel Conversion:} All audio recordings were resampled to a common sampling rate of 44.1 kHz and converted to mono channel, 16-bit resolution. This ensures consistency across all input samples.
    \item \textbf{Log Mel Spectrograms:} The primary features extracted were log-scaled mel spectrograms. These time-frequency representations are widely used in audio processing and have proven effective for various sound event detection tasks. They were computed using 128 mel bands, a window size of 23 milliseconds (ms), and a hop size of 11.5 ms. The mel scale mimics human auditory perception, making these features more perceptually relevant. \cite{10059143}
    \item \textbf{Auxiliary Features:} In addition to mel spectrograms, we also explored other features for potential fusion or alternative classification approaches:
        \begin{itemize}[leftmargin=*]
           \item \textbf{Autocorrelation of Waveform:} This feature can capture periodicity and temporal patterns in the raw audio signal, which might be indicative of certain gunshot characteristics (e.g., echoes, rapid fire).
          \item \textbf{Bag of Audio Words (BoAW):} For classifier fusion approaches, a BoAW representation was considered. This involves creating a dictionary of acoustic "words" (e.g., using k-means clustering on low-level descriptors) and representing audio segments as histograms of these words. \cite{schmitt2017openxbow, schmitt2016border,Shah2025,10.1145/2964284.2964310}
        \end{itemize}
\end{enumerate}
For the primary experiments reported in this paper, log mel spectrograms served as the input to our deep learning model.

\subsection{Classification Models}
\label{ssec:classifiers}
We evaluated two main types of classifiers: a traditional machine learning model (SVM) to establish a baseline, and a deep learning model (CNN) for potentially improved performance. We also implemented Audio Spectrogram Transformers \cite{gong2021ast}, however it was challenging in this scenario due to the amount of available data. 

\subsubsection{Support Vector Machine (SVM) Baseline}
A Support Vector Machine (SVM) was employed as a baseline classifier. SVMs are powerful supervised learning models that find an optimal hyperplane to separate data points belonging to different classes. For this experiment, features (likely global statistics derived from spectrograms or BoAW representations) were fed into an SVM configured for multi-class classification.

\subsubsection{Convolutional Neural Network (CNN) for Joint Detection and Classification}
\begin{figure}[h!]
  \centering
  \includegraphics[width=\linewidth]{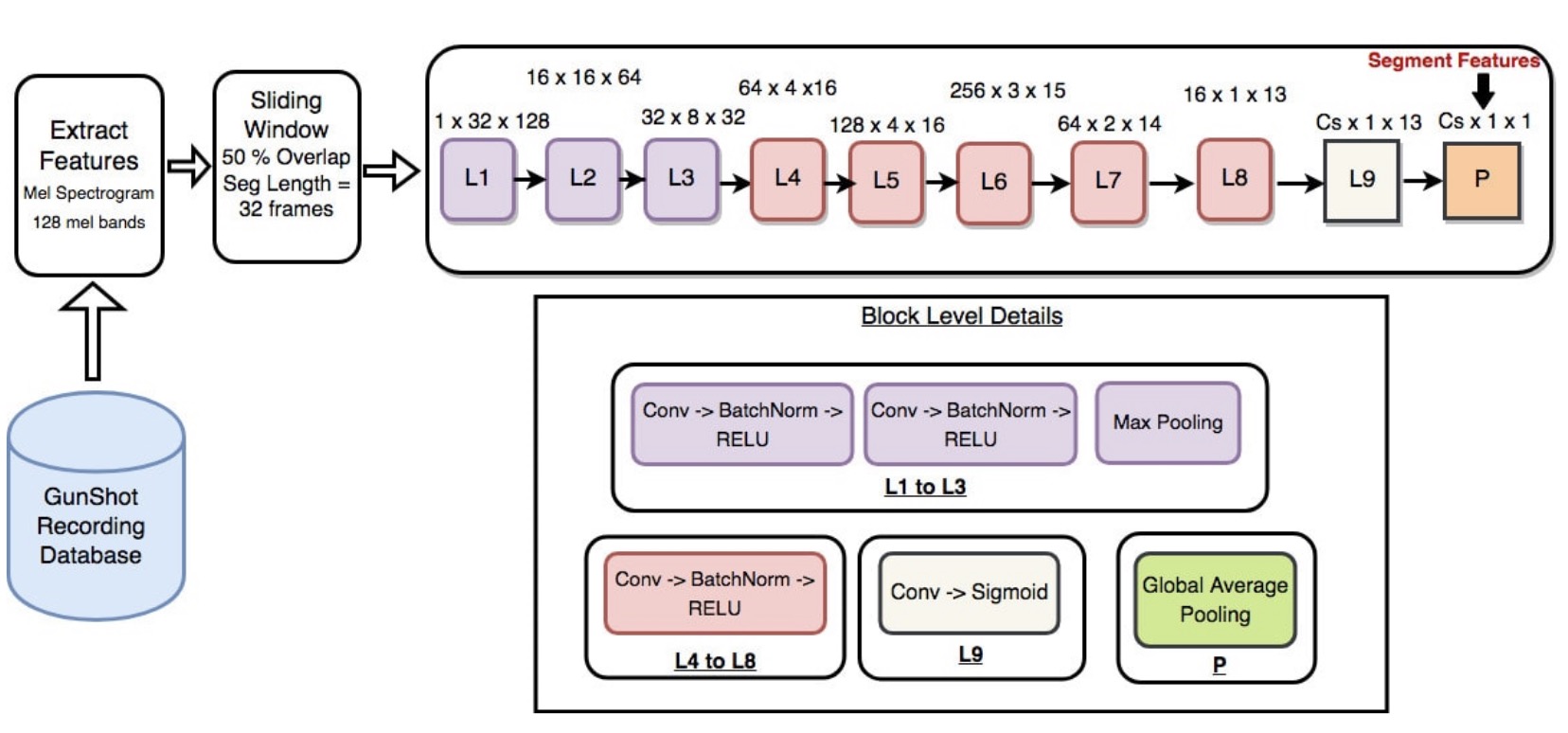} 
  \caption{Joint Detection and GunType Classification CNN Architecture. The network takes audio features (e.g., mel spectrograms) as input and has shared convolutional layers followed by separate branches for gunshot detection and gun type classification.\cite{Shah2025}}
  \label{fig:joint_detection_architecture}
  \vspace{-2mm}
\end{figure}

To leverage the spatial and temporal structure inherent in spectrograms, a Convolutional Neural Network (CNN) was designed. CNNs have demonstrated state-of-the-art performance in various audio and image recognition tasks. Our architecture, depicted in Figure \ref{fig:joint_detection_architecture}, is designed for joint learning: it simultaneously performs gunshot detection (binary classification: gunshot vs. no-gunshot) and gun type classification (multi-class classification among the five gun types). \cite{han2018co}

The architecture typically involves:
\begin{itemize}
    \item \textbf{Input Layer:} Accepts the log mel spectrograms.
    \item \textbf{Convolutional Layers:} A series of convolutional layers with activation functions (e.g., ReLU) and pooling layers (e.g., max pooling) to extract hierarchical features from the spectrogram. These layers learn to identify patterns like edges, textures, and eventually more complex acoustic events.
    \item \textbf{Shared Layers:} Early convolutional layers are often shared between the two tasks, allowing the model to learn common low-level features relevant for both detecting a gunshot and identifying its type.
    \item \textbf{Task-Specific Branches:} After the shared layers, the network splits into two branches:
        \begin{itemize}
            \item \textbf{Gunshot Detection Branch:} Typically ends with a sigmoid activation function for binary classification.
            \item \textbf{Gun Type Classification Branch:} Ends with a softmax activation function for multi-class classification. This branch is only activated or considered if the detection branch positively identifies a gunshot.
        \end{itemize}
    \item \textbf{Fully Connected Layers:} Dense layers are usually present before the output layers in each branch to perform high-level reasoning.
\end{itemize}
This joint learning approach can be beneficial as the features learned for gunshot detection can aid in gun type classification, and vice-versa.

\subsection{Training and Evaluation Protocol}
\label{ssec:training_protocol}
The curated dataset (Table \ref{tab:dataset_summary}) was partitioned into training, validation, and testing sets to ensure robust model evaluation:
\begin{itemize}[leftmargin=*]
    \item \textbf{Training Set (60\%):} Used to train the parameters of the models.
    \item \textbf{Validation Set (20\%):} Used to tune hyperparameters and for early stopping to prevent overfitting during training.
    \item \textbf{Testing Set (20\%):} Used for the final evaluation of the trained models on unseen data.
\end{itemize}
To further assess the robustness and generalization capability of the architectures, \textbf{5-fold cross-validation} was employed on the training/validation partitions. The primary evaluation metrics used are Precision, Recall, and F1 Score, calculated for each class and overall. Mean Average Precision (mAP) is also reported as a summary statistic for the gun type classification task.

i

\subsection{Experiments and their design}
\label{ssec:experiments}
Two main experiments were designed:

\textbf{Experiment 1: Baseline performance with SVM}
This experiment aimed to establish a baseline performance for gun-type classification using a traditional SVM classifier \cite{suthaharan2016support}. The features used for the SVM were likely derived from the audio representations described in Section \ref{ssec:features}.

\textbf{Experiment 2: Joint Gunshot Detection and GunType Classification with CNN}
This experiment involved training and evaluating the proposed CNN architecture (Figure \ref{fig:joint_detection_architecture}). The model was trained to simultaneously predict: Whether an audio segment contains a gunshot and If a gunshot is present, which of the five gun types it belongs to.
Performance was evaluated on the clean, annotated dataset. Additionally, the performance of models trained or tested using noisy data sourced from the web (e.g., YouTube soundtracks) was investigated to understand real-world generalization capabilities. \cite{Shah2025}

\section{Results and Discussion}
\label{sec:results}
This section presents the performance of the classification models on the tasks of gunshot detection and gun type classification. We analyze the results obtained from both the SVM baseline and the deep learning CNN architecture.

\subsection{Gunshot Detection Performance}
\label{ssec:gunshot_detection_results}
The first task for the joint learning model is to accurately detect the presence of a gunshot in an audio segment. Table \ref{tab:gunshot_detection_results} summarizes the precision, recall, and F1 score for the binary classification task of "Gunshot" vs. "No Gunshot" using the CNN model.

\begin{table}[h!]
  \centering
  \caption{Performance of Gunshot Detection (CNN Model).}
  \label{tab:gunshot_detection_results}
  \resizebox{0.7\columnwidth}{!}{%
  \begin{tabular}{@{}lccc@{}}
    \toprule
    \textbf{Class} & \textbf{Precision} & \textbf{Recall} & \textbf{F1 Score} \\
    \midrule
    No Gunshot     & 0.96               & 0.86            & 0.90              \\
    Gunshot        & 0.82               & 0.91            & 0.86              \\
    \bottomrule
  \end{tabular}%
  }
\end{table}

The model demonstrates strong performance in detecting gunshots, achieving an F1 score of 0.86 for the "Gunshot" class and 0.90 for the "No Gunshot" class. The high recall of 0.91 for "Gunshot" indicates that the system is effective at identifying most actual gunshot events, which is crucial for a system intended to alert law enforcement. The precision of 0.82 for "Gunshot" means that when the system flags an event as a gunshot, it is correct 82

\subsection{Gun Type Classification Performance}
\label{ssec:guntype_classification_results}
For the audio segments correctly identified as containing a gunshot, the second task is to classify the type of firearm. Table \ref{tab:guntype_classification_results} presents the detailed per-class precision, recall, and F1 score for the gun type classification task using the CNN model. The table includes two sets of metrics: overall metrics and "Relevant" metrics. The "Relevant" metrics likely refer to performance calculated only on instances that were correctly detected as gunshots by the first stage of the joint model, or perhaps after applying a specific confidence threshold. \cite{shah2024importance} For this discussion, we will assume "Relevant" metrics are conditioned on correct gunshot detection.


\begin{table}[ht!]
  \centering
  \caption{Performance of Gun Type Classification (CNN Model). "Relevant" scores are conditioned on correct prior gunshot detection.}
  \label{tab:guntype_classification_results}
  \small 
  \resizebox{0.95\columnwidth}{!}{%
  \begin{tabular}{@{\hskip 2pt}lcccccc@{\hskip 2pt}}
    \toprule
    \multirow{2}{*}{\textbf{Class}} & \multicolumn{2}{c}{\textbf{Precision}} & \multicolumn{2}{c}{\textbf{Recall}} & \multicolumn{2}{c}{\textbf{F1 Score}} \\
    \cmidrule(lr){2-3} \cmidrule(lr){4-5} \cmidrule(lr){6-7}
    & Overall & Relevant & Overall & Relevant & Overall & Relevant \\
    \midrule
    Rifle          & 0.21 & 0.49 & 0.44 & 0.52 & 0.28 & 0.50 \\
    Submachine Gun & 0.53 & 0.76 & 0.43 & 0.87 & 0.48 & 0.81 \\
    Handgun/Pistol & 0.59 & 0.70 & 0.59 & 0.58 & 0.59 & 0.64 \\
    Machine Gun    & 0.75 & 0.79 & 0.51 & 0.78 & 0.61 & 0.78 \\
    Shotgun        & 0.30 & 0.386 & 0.28 & 0.40 & 0.29 & 0.39 \\
    \bottomrule
  \end{tabular}%
  }
\end{table}

Observing the "Relevant" F1 scores, which provide a clearer picture of the gun type classification capability itself:
\begin{itemize}[leftmargin=*]
    \item \textbf{Submachine Gun} and \textbf{Machine Gun} classes achieve the highest F1 scores (0.81 and 0.78, respectively). This suggests their acoustic signatures (e.g., rapid fire rate, specific timbre) are relatively distinct and well-learned by the model. Machine Guns show high precision (0.79 relevant), indicating strong confidence when this class is predicted. Submachine Guns show excellent relevant recall (0.87), meaning the model is good at finding most instances of this class.
    \item \textbf{Handgun/Pistol} achieves a moderate F1 score (0.54 relevant). This category is broad and can encompass a wide variety of calibers and mechanisms, potentially leading to more acoustic variability and confusion with other classes.
    \item \textbf{Rifle} also shows a moderate F1 score (0.50 relevant). Similar to handguns, rifles span a range of actions and calibers. The relatively lower precision (0.49 relevant) suggests it might be confused with other long guns or even some powerful handguns.
    \item \textbf{Shotgun} exhibits the lowest F1 score (0.39 relevant). This could be due to fewer training samples (as seen in Table \ref{tab:dataset_summary}) or perhaps its distinctive but singular report being occasionally confused with other loud, impulsive sounds or even single shots from other firearm types under noisy conditions.
\end{itemize}
The "Overall" scores (which likely factor in errors from the initial gunshot detection phase or are calculated across all data without pre-filtering) are consistently lower than the "Relevant" scores, highlighting the importance of accurate initial gunshot detection for the subsequent classification task.

\subsection{Comparative Performance and Impact of Data Quality}
\label{ssec:comparative_performance}
A key finding is the comparison between the traditional machine learning approach and the deep learning model, as well as the impact of data quality:
\begin{itemize}[leftmargin=*]
    \item \textbf{SVM vs. CNN on Clean Data:} Using the clean, labeled dataset for training and testing, the SVM classifier achieved a Mean Average Precision (mAP) of \textbf{0.39} for gun type classification. In contrast, the proposed Deep Learning (CNN) architecture achieved a significantly higher mAP of \textbf{0.58}. This demonstrates the superior capability of CNNs to learn complex patterns from spectrogram data for this task.
    \item \textbf{Performance with Noisy Web Data:} When the system was evaluated using noisy data sourced from the web (e.g., YouTube videos, which often contain background noise, music, and variable recording quality), the mAP dropped to \textbf{0.35}. This highlights a critical challenge: generalizing models trained on relatively clean, curated data to the diverse and often poor-quality audio encountered in real-world, uncontrolled environments \cite{kumar2018learning}. The "domain shift" between clean training data and noisy web data significantly impacts performance.\cite{shah2018closer}
\end{itemize}

The results indicate that while promising performance can be achieved on curated datasets, significant work is needed to improve robustness against noise and varying acoustic conditions typically found in "in-the-wild" recordings. The observation that "full audio recording extracted features performs poorly for Gun type classification" likely refers to using global features from entire, potentially long and noisy recordings without precise temporal localization of the gunshot event, which the joint detection-classification CNN aims to address by focusing on relevant segments.

\section{Conclusions}
\label{sec:conclusions}

This research demonstrates the feasibility of using commodity hardware for critical public safety applications, potentially democratizing access to gunshot detection technology. Our key findings show that while deep learning approaches significantly outperform traditional methods on curated data, the transition to real-world deployment remains challenging.  We curated a dataset of 3459 recordings and investigated the fundamental acoustic properties of gunshots that enable such classification. Our primary contributions and findings are:


\begin{enumerate}
    \item We demonstrated that a Convolutional Neural Network (CNN) designed for joint gunshot detection and gun type classification significantly outperforms a traditional Support Vector Machine (SVM) baseline on clean, annotated data. The CNN achieved a Mean Average Precision (mAP) of 0.58, compared to 0.39 for the SVM.
    \item The CNN model showed strong performance in binary gunshot detection (F1 score of 0.86 for gunshots) and achieved promising, albeit varied, results for classifying specific gun types, with Submachine Guns and Machine Guns being identified most reliably.
    \item A significant challenge lies in generalizing models to noisy, real-world data. Performance dropped considerably (mAP of 0.35) when evaluated on data sourced from the Web, underscoring the need for robust feature extraction, noise-resilient models, and domain adaptation techniques. 
    \item The study confirms that classifying gun types based on acoustic signatures is a complex problem, influenced by data quality, environmental factors, and the inherent acoustic similarity between certain firearm categories.
\end{enumerate}

While the results are promising, particularly with the deep learning approach on curated data, the path to a highly reliable system deployable in diverse real-world scenarios requires further research.

\section{Future work}
\label{sec:future_work}

Building upon current findings, ongoing and future research addresses limitations and advances towards a practical system. Immediate focus is on improving model robustness and accuracy by: \textbf{Addressing Label Noise} (exploring methods to account for and correct incorrect or ambiguous labels in training data); and \textbf{Advanced Data Augmentation and Synthesis} (investigating sophisticated techniques to improve generalization and handle data scarcity) \cite{zhang2017mixup}. The overarching goals extend beyond the current study: \textbf{Cost-Effective Deployment} (dramatically reducing system costs to enable operation on readily available hardware) \cite{morehead2019low}; \textbf{Real-Time, Highly Accurate Alerts} (developing a system for timely, accurate alerts from various audio sources); and \textbf{Shooting Location Estimation} (exploring methods to estimate shooting location relative to recording devices) \cite{10.1145/3343031.3350536}. The achievement of this vision aims to improve public safety with accurate, timely information while increasing accessibility.

\clearpage
\bibliographystyle{IEEEtran}
\bibliography{refs25}







\end{document}